# Single-pump hybrid nonlinearities in transparent conductors

Wallace Jaffray[1], Sven Stengel[1], Alexandra Boltasseva[2], Vladimir M. Shalaev[2], Carlo Rizza[3], Domenico de Ceglia[4], Maria Antonietta Vincenti[4], Michael Scalora[4], Matteo Clerici[5], and Marcello Ferrera[1,*]

[1]Institute of Photonics and Quantum Sciences, Heriot-Watt University, Edinburgh, UK
[2]Elmore Family School of Electrical & Computer Engineering and Birck Nanotechnology Center, Purdue University, West Lafayette, Indiana 47907, USA
[3]Department of Physical and Chemical Sciences, University of L'Aquila, L'Aquila, Italy
[4]Department of Information Engineering,University of Brescia,Brescia, Italy
[5]Department of Science and Technology, University of Insubria, Como, Italy
[*]Email: m.ferrera@hw.ac.uk

**Abstract**

Low-index transparent conducting oxides have attracted significant attention because ultrafast optical excitation in these materials can induce exceptionally large temporal index gradients. Due to this remarkable nonlinear optical behaviour, this material platform enables sub-picosecond, all-optical control of photon energy and momentum, with growing relevance for integrated photonics, quantum optics, and optical computation. Owing to their hybrid electronic structure, transparent conductors exhibit both intraband and interband nonlinearities, previously accessed using dual-colour excitation with near-infrared and ultraviolet pumps. Here, we show that both excitation regimes can be activated using a single, intense near-infrared pump. Above a threshold intensity, the pump drives hot-electron intraband dynamics while simultaneously generating higher harmonics that trigger interband excitation. The interplay of these two effects sharpens the temporal features of the recorded transmissivity which in turn substantially broadens the effective material bandwidth. Finally, by comparing linear and circular pumping conditions, we further demonstrate that the observed interband nonlinearities originate from harmonic generation rather than from direct multiphoton absorption. Our results provide key insights into the strong-field optical response in these time-varying photonic materials, opening new frontiers for the ultra-fast manipulation of photons in both classic and quantum regimes.

## Introduction

Traditionally, the momentum of light can be controlled by spatially structuring the refractive index. This is the case for lenses, gratings, and more recently metamaterials [1]. However, purely spatial designs cannot access all regimes of light–matter interaction and can only control photon momentum. In principle, temporally engineering the refractive index of materials allows for exerting control over both momentum and energy of propagating photons [2, 3]. However, modifying material properties by large amounts and within optical timescales is a challenging task.

When targeting very large refractive index time gradients, plasma nonlinearities have long been considered the closest approximation to time-varying systems. They are rooted several physical mechanisms, such as modifications of plasma density [4] or relativistic effects [5], and can create strong changes in the refractive index over timescales comparable with the optical cycle. Thus they can shift the frequency of incoming light by an appreciable margin [3, 6, 7]. However, despite the strong nonlinearities available in plasmas, a solid-state based approach would be greatly desirable due to critical technological advantages in terms of compactness and usability [8].

When moving from plasma to solid state physics, transparent conducting oxides (TCOs) are considered an intermediate step given their hybrid nature, as they exhibit properties of both metals and semiconductors. A key distinction between TCOs and true plasmas lies in the presence of a crystal lattice that under certain conditions (e.g., high field intensity) actively participates in shaping the overall optical properties of the system. Within this context, metals are not considered as an optimal alternative for nonlinear optics applications given their poor tunability, weak coupling to light, and low damage thresholds [9].

As previously stated, TCOs are hybrid in their nature, originating from wide-bandgap compounds which, through tailored fabrication and doping processes, achieve intrinsically high free-carrier concentrations typically only one order of magnitude lower than those of metals [10–13]. TCOs can exhibit a low index window at near-infrared wavelengths, which can enhance nonlinearities via slow-light and other effects related to the near-zero-index reigeme [14]. From a microscopic point of view, the hydrodynamic free-electron response, analogous to those in metals and plasmas, is responsible for substantial nonlinearities including non-centrosymmetric contributions that enable even-order harmonic generation despite the material's bulk centrosymmetry [15, 16]. The most accessible nonlinear mechanism however, pertains intraband absorption and the heating of conduction band electrons, which activates at very low optical excitation energies. More specifically, due to the non-parabolic nature of TCO's conduction band, carrier temperature is tied to the carrier effective mass [17], which modulates the plasma frequency $\omega_p$, and drives large refractive index changes (on the order of unity) [18, 19] over ultrafast timescales ($< 10$ fs) [20].

Additionally, unlike plasmas, where the number of free electrons is fixed, in solids the electron population can be dynamically altered through optical excitation from bound states. In this direction, Clerici et al. [21] demonstrated that dual-pump excitation at carefully chosen wavelengths can independently access two distinct nonlinear pathways in TCOs: interband transitions that increase carrier concentration and thus blueshift the ENZ wavelength, and intraband hot-electron heating that increases the effective mass and thus redshifts the ENZ wavelength. This bidirectional control enables unprecedented flexibility in modulating the refractive index, with demonstrations showing bandwidth-large frequency shifts while also achieving internal conversion efficiencies greater than unity through time-refraction processes [22].

In this work, by leveraging both metallic nonlinearities and the absorption of generated harmonics below the bandgap, we demonstrate both positive and negative ultrafast refractive index change using a single-pump excitation at a wavelength of 787 nm. In addition, by tuning the probe wavelength across the low index spectral window, we show that the relative strength of positive and negative index change can be tuned. Our results also establish a direct and effective route to temporally enlarge the effective material bandwidth. This capability of engineering the time-varying optical responses enables practical implementation in ultrafast photonic devices, frequency-division multiplexing architectures, and photonic time crystals. This work also advances the fundamental understanding of the underlying material physics by elucidating the interplay between free- and bound-electron nonlinearities under high-intensity optical excitation.

## Experimental methods

Experiments were performed using a standard pump–probe configuration operating close to normal incidence (Fig. 1). Pump pulses have their wavelength centred at 787 nm, with a full width half maximum (FWHM) duration of 100 fs. The pump was then focused onto a spot size with diameter of 1 mm on the sample. The probe beam, which has FWHM duration of 85 fs, was attenuated by several orders of magnitude to avoid self-action nonlinearities, and it was focused to a spot size of $\approx 200$ µm. Both pump and probe polarizations were set parallel to the optical table and aligned within the plane of incidence. Transient transmissivity measurements were taken by changing the time delay between pump and probe

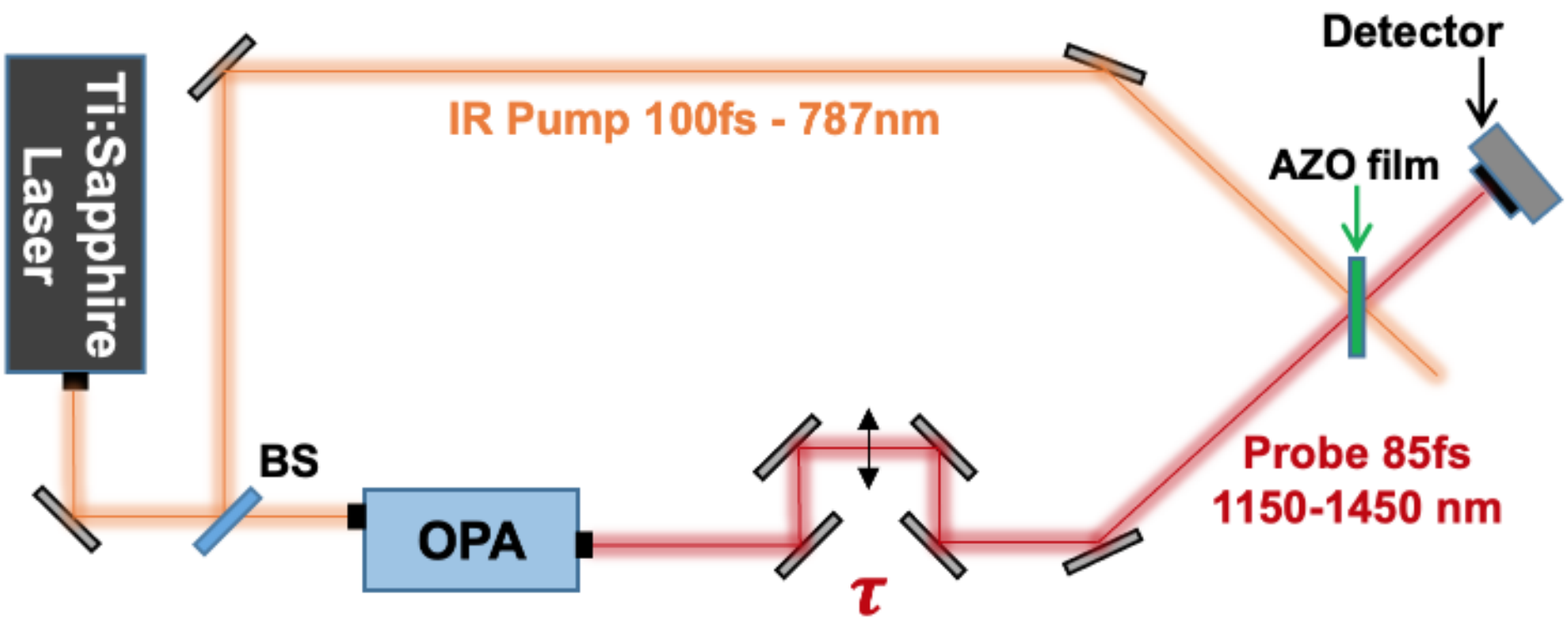

Figure 1: **Experimental setup.** Pump-probe setup using 100 fs pump pulses at 787 nm and 85 fs probe pulses with central wavelength tunable between 1150 nm and 1450 nm. Transmitted light is captured by a calibrated near-infrared photodiode. Both pump and probe are at almost normal incidence despite the pictorial representation, which exaggerates angle for ease of viewing. The experiments were conducted with the pump and probe beams incident at near-normal incidence ($< 5°$).

in steps of a few femtoseconds.

All measurements were carried out on a 900-nm-thick aluminium-doped zinc oxide (AZO) film, where we monitored the transient transmission of the probe under optical pumping. In the first configuration, the probe wavelength was set to 1350 nm (near the AZO's ENZ crossover), and the pump power was finely scanned from 100 GW/cm$^2$ up to 2400 GW/cm$^2$. This is still considerably below the damage threshold of the materials [23]. Next, probe pulses were tuned to specific wavelengths across the low-index window 1150 nm, 1250 nm, 1350 nm, and 1450 nm. For each of these probe wavelengths, pump-probe experiments were conducted at different pump intensities corresponding to 435, 869, and 1304 GW/cm$^2$. Finally, to verify the interband mechanism, the pump was set to a circular polarisation and a pump-probe scan was completed at 1150 nm with an intensity of 1304 GW/cm$^2$. Further details on this measurement are reported in the subsequent paragraph.

## Results and discussion

The targeted experimental parameter of interest was the transient relative probe transmission as a function of both the pump-probe delay $\tau$ and the pump intensity $I$. This quantity is defined as $\Delta T(\tau, I) = (T(\tau, I) - T_0)/T_0$, where $T_0$ is the linear transmission at the probe wavelength and $T$ is the instantaneous probe transmission. Figure 2a shows experimentally attained values in the color map. An overview of simulation results, attained for identical experimental parameters, is shown in the inset of Fig. 2a. Here the relative transmission modulation is computed via full-wave simulations where Maxwell equations are coupled with auxiliary equations describing the material response (see Theory Section for further details).

A very critical and central part of our study concerns the change in behaviour of $\Delta T$ that happens when the pump fluences is increased. To properly highlight this material evolution in the transmission time profile, Fig. 2b shows two intensity cut-lines at low and high intensities ($I$ = 494 GW/cm$^2$, blue circles and $I$ = 2306 GW/cm$^2$, orange circles) taken from Fig. 2a. Related numerical estimates using our model are also reported in solid and dashed black lines, respectively. At low pump intensities $\Delta T$ undergoes a positive change as pump and probe overlap in time, and it returns to the linear state within a few hundred femtoseconds, as expected for the intraband hot-electron nonlinearity. As the intensity is increased beyond $\approx 1$ TW/cm$^2$, the pump-probe delay $\tau$ at which $\Delta T$ reaches its maximum, shifts to earlier delays. Additionally the temporal width of the hot-electron nonlinear response begins to "compress"; a point which we will discuss in more detail below. Finally, at intensities just below 2 TW/cm$^2$, a negative transmission modulation is observed, which is comparable in magnitude to the hot-electron nonlinearity. This effect can be attributed to TCOs interband nonlinearity, which was previously investigated in [21] via a hybrid UV and NIR pumping scheme. However, as our setup has only a single pump at 787 nm, the interband nonlinearity is instead driven via harmonic generation below the band edge. This claim will be experimentally verified at the end of this section.

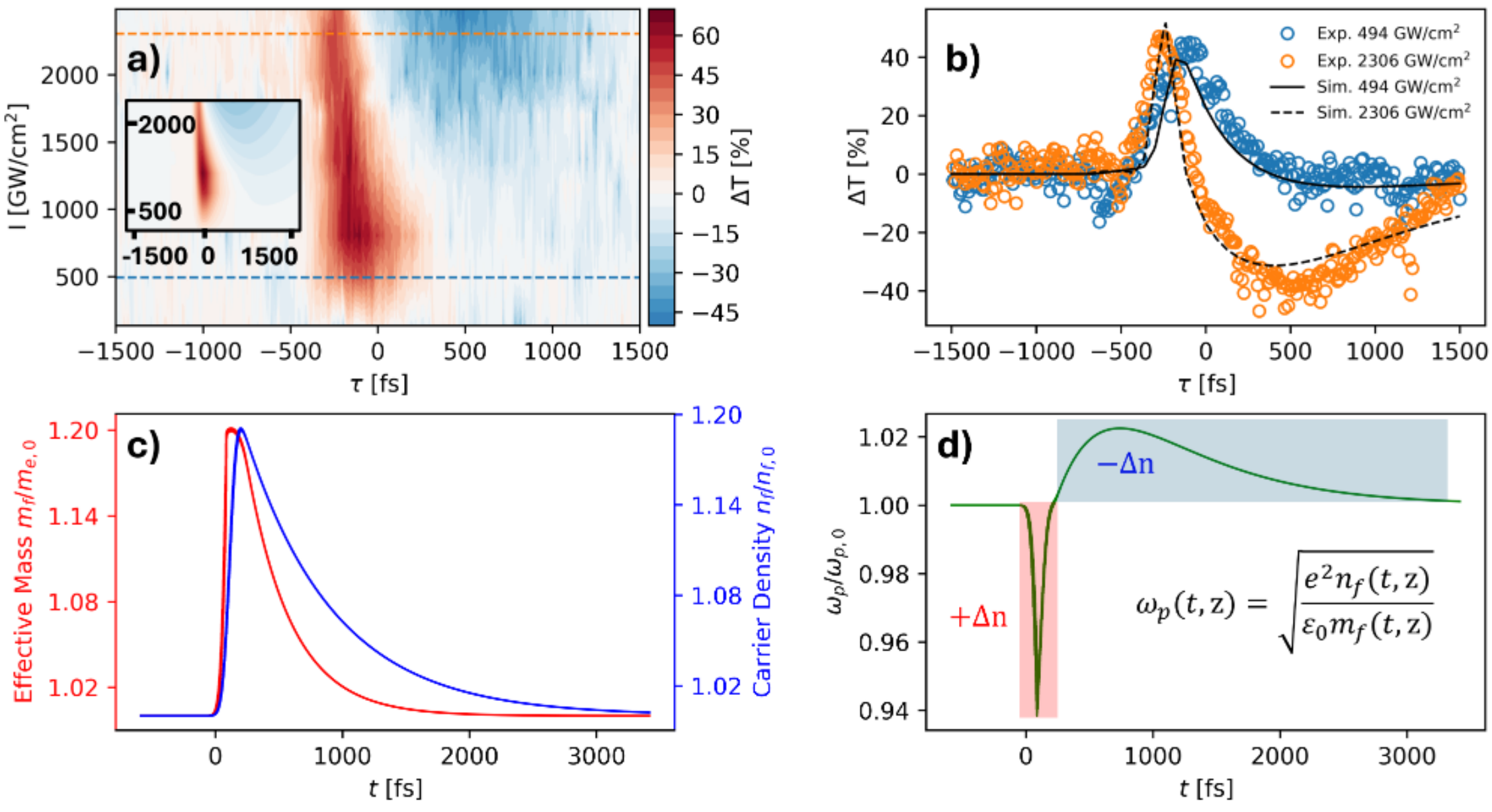

Figure 2: **Pump-probe study as a function of pump intensity.** a) Heatmap of probe relative transmission (defined as $\Delta T(\tau, I) = (T(\tau, I) - T_0)/T_0$, where $T_0$ is the linear transmission) as a function of pump-probe delay $\tau$ and pump intensity $I$. The pump and probe wavelengths were fixed at 787 nm and 1350 nm, respectively. b) Cut-lines plots for pump intensities of 494 GW/cm$^2$ (blue line) and 2306 GW/cm$^2$ (orange line). c) Dual axis plot of the electron effective mass (left axis and red line) and carrier concentration (right axis and blue line) as a function of simulation time $t$ for a pump intensity of 2306 GW/cm$^2$. Values are normalised with the electron rest mass $m_{e,0}$ and the rest free carrier density $n_{f,0}$. d) Instantaneous plasma frequency $\omega_p(t, z)$ as modulated by the pump wave. The positive and negative index changes are indicated by red and blue highlighted sections, respectively.

To understand the complex evolution of the material's response for very high pump intensities (Fig. 2b) we focus on two key parameters regulating the intraband and interband nonlinearities. These can be easily identified in the optically induced temporal change of the effective mass (red line) and carrier density (blue line), respectively, as plotted in Fig. 2c. The interplay of these two parameters provides the overall nonlinear response of the material, which is regulated by the time-varying plasma frequency $\omega_p(t, z)$ (Fig. 2d). This parameter drives the time-varying index of the material as is evident from comparing the shape of high-power transmission modulation (orange line Fig. 2b) and the time-varying plasma frequency (green line Fig. 2d). Right after the pump-pulse strikes the film, the plasma frequency sharply decreases due to carrier heating. Given a short delay, the interband process repopulates the lower levels of the conduction band and overcomes the hot-electron process, causing the plasma frequency to rise. This interaction allows for creating a sharp modulation of the plasma frequency with FWHM duration of 50 fs by using a 100 fs pump pulse. It is worth underlining that under high intensity pumping, the temporal features of the material's response become considerably shorter than the optical excitation. This leads to a dramatic enlargement of the material bandwidth, which could find applications in THz science and for the generation of photonic time crystals [24]. It is also worth discussing the apparent temporal shift of transmissivity peak, which appears to happen at slightly earlier times for increasing intensities (see Fig.2a). This is due to the increase in carrier density (interband nonlinearities) acting against the increase of carrier effective mass (intraband nonlinearities) at lower intensities (Fig. 2c).

To investigate the wavelength dependence of these nonlinearities, we performed pump-probe measurements across the low-index spectral window of the sample. Figures 3a–d show the experimentally measured transient transmission $T(\tau, I)$ for probe wavelengths $\lambda_p$ of 1150 nm, 1250 nm, 1350 nm, and 1450 nm, respectively. For each probe wavelength, the pump power was varied to 435, 869, and 1304 GW/cm$^2$. In Fig. 3a, the probe transmission shows a pronounced initial positive transmission change even at low pump powers, which is then followed by the emergence of negative modulation at higher delays. This indicates that at probe wavelengths closer to the UV transition, the strength of the interband nonlinearity increases. This trend is evident for 1250 nm and 1350 nm, (Fig. 3b and c), where the interband nonlinearity clearly diminishes further, and finally, at 1450 nm (Fig. 3d), only the intraband effects are observable. Figures 3e–h present corresponding simulations which use the same parameters as used in Fig. 2. The diminishing behaviour of the interband nonlinearity is captured by a wavelength dependant interband coupling parameter $\Lambda(\lambda_p)$, the values of which can be found in Table 1.

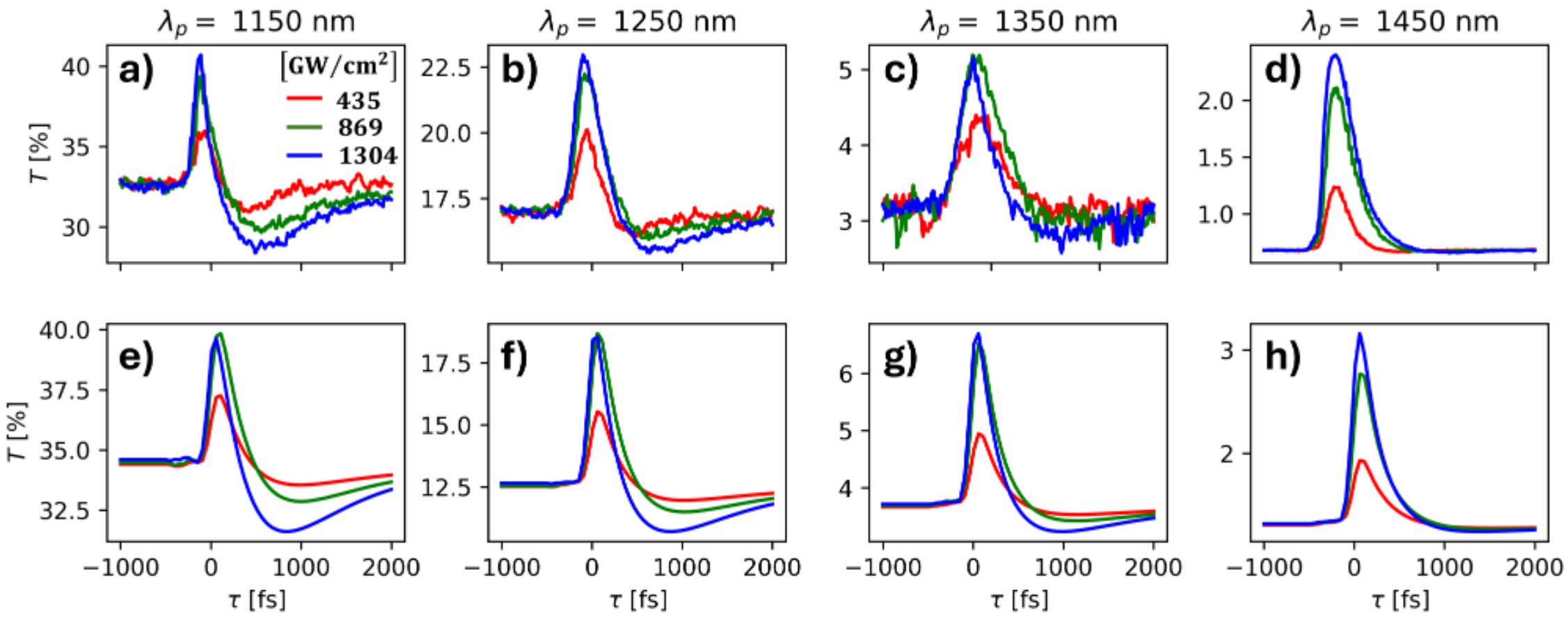

Figure 3: **Probe wavelength scan.** a), b), c), and d) Provide the probe transmission $T(\tau, I)$ as a function of the pump delay $\tau$ and pump power $I$, for probe wavelengths of 1150 nm, 1250 nm, 1350 nm, and 1450 nm. The pump power was varied to 435, 869, and 1304 GW/cm$^2$. e), f), g), and h) show the theoretical predictions corresponding to cases (a–d), respectively.

A key distinction of our work from previous studies is that the interband effect are driven by highly efficient harmonic generation in TCOs [16, 25, 26]. Specifically, the 787 nm pump efficiently generates harmonics which reside below the band edge, exciting the UV-transition without the need for a dedicated secondary pump [16]. However, one might question whether the observed change in free-electron density is driven predominantly by nonlinear multiphoton absorption, rather than by linear absorption of higher-frequency components generated through harmonic conversion.

To verify which of these interband mechanisms is dominant, we set the pump to circular polarisation which largely disables the generation of odd harmonics while keeping multiphoton absorption unaffected [27]. Although harmonic suppression is absolute under ideal conditions, small deviations from the ideal case (e.g. residual anisotropy, finite bandwidth, or spatial field inhomogeneity) can still allow weak harmonic generation, which is expected to be negligible under our experimental conditions [28, 29]. For the sake of completeness, it is also worth mentioning that in systems with strong lattice anisotropy or complex band-structure, such as crystalline metals, circularly polarised excitation can still produce relevant odd harmonics. Additionally, while even order harmonics will continue to be generated in our film, they are not strong enough to activate the interband nonlinearities we observe [16]. Figure 4 compares the transient transmission of a 1150 nm probe with circular and linearly polarised pumping at 1300 GW/cm$^2$. As expected, linear pumping produces both a positive and negative transmission change (blue line) while circularly polarised pumping (red line) only exhibits an increase in transmission, associated with the hot electron nonlinearity. In the latter case, the interband nonlinearity has been completely suppressed as there are no generated odd harmonics to pump the UV transition. One may notice that the transmission increase (intraband effect) is also reduced for the circular pumping case; this is a well-known effect which acts to reduce the strength of nonlinearities induced by circular polarised pumping when compared to linearly polarised pumping [30].

Finally, using our prior knowledge on harmonic generation efficiencies in AZO films [16], we can estimate the reabsorbed harmonic photon flux and identify the specific harmonic channel which excites the interband transition. Specifically, we focus on the third harmonic (TH) and fifth harmonic (FH) emissions from our 787 nm pump beam at around 1 TW/cm$^2$. Considering a 900 nm film and 100 fs pulse, our internal TH intensity is around 100 MW/cm$^2$ with wavelength of 262 nm [16]. The number of TH photons per pulse $N_{3\omega}$ can be approximated as follows:

$$N_{3\omega} = \frac{I_{3\omega}}{3\hbar\omega}\Delta\tau \approx 1.3 \times 10^{13} \text{ photons/cm}^2 \tag{1}$$

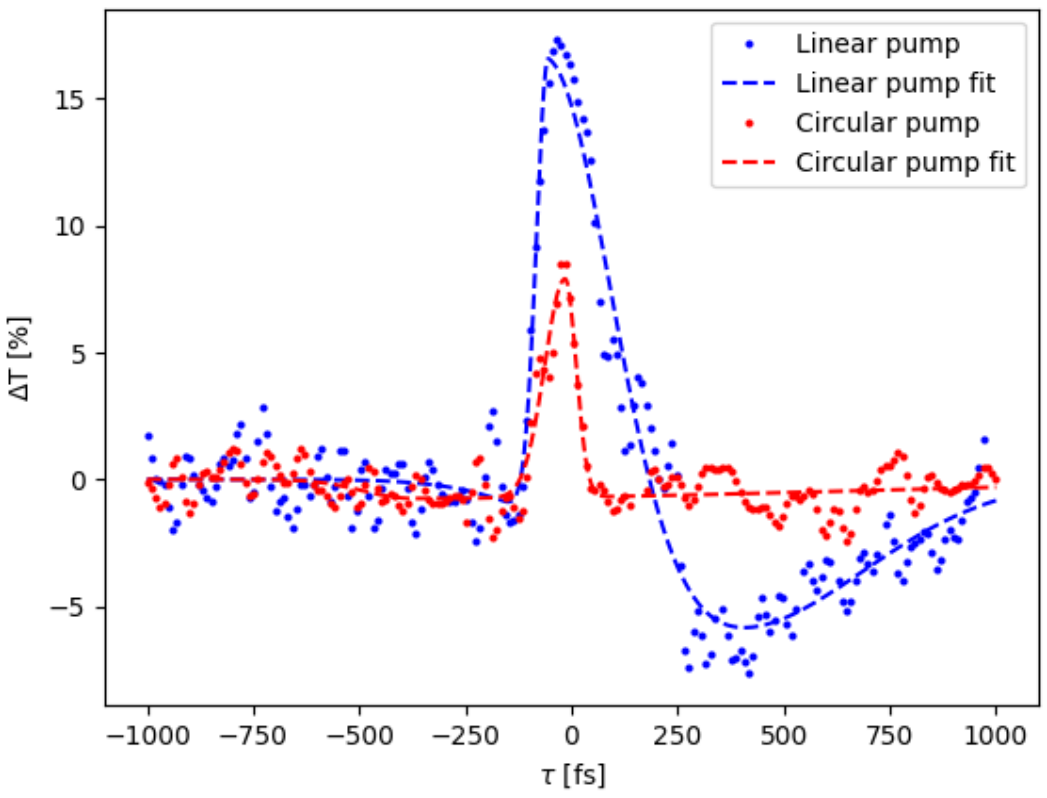

Figure 4: **Circular versus linear pump-probe.** Relative transmission change for a 85 fs probe centered at 1150 nm, pumped with linearly (blue line and dots) and circularly (red line and dots) polarised pulses at 1300 GW/cm$^2$. The latter pumping case eliminates the generation of odd harmonics, and consequently suppresses the interband effect.

where $\Delta\tau$ is the pulse duration, $\omega$ is the fundamental frequency, and $I_{3\omega}$ is the TH intensity. Assuming all TH photons are absorbed by the film, we find the relative carrier density change $\Delta n_f^{3\omega}$:

$$\Delta n_f^{3\omega} \approx \frac{N_{3\omega}}{n_{f,0} d} \approx 3 \times 10^{-2}\% \tag{2}$$

where $d$ is the film thickness and $n_{f,0}$ is the unpumped free carrier density. As this value is much smaller than the $\approx$ 10% change suggested by the model, we repeat the same calculation for the FH. Local field simulations indicate that the internal FH intensity can reach 5 GW/cm$^2$ (at 167 nm) considering a 1 TW/cm$^2$ pump. While this value may seem high, it is worth stressing that this is an internal intensity and the emitted FH is many orders of magnitude lower, as would be expected from previous measurements. Using the same process as before we now find the free carrier density change to be $\Delta n_f^{5\omega} \approx 9\%$. This aligns closely with the measured results, and identifies the FH generation as the dominant channel for activating the interband nonlinearity.

# Theoretical analysis

Let us consider 1+1D pump and probe scenario, where both pulse trains impinge normally on an AZO slab and propagate along the z-axis. To fully capture experimentally recorded effects, we consider the dynamics of the time dependant conduction band electron density $n_f(t,z)$ (modelling our interband process) and the conduction band electron effective mass $m_f(t,z)$ (modelling hot-carrier generation). From these parameters, it is possible to deduce the following equation which captures both measured nonlinearities [15]:

$$\frac{\partial^2 P_f}{\partial t^2} + \left(\gamma_f + \frac{1}{m_f(t,z)}\frac{\partial m_f(t,z)}{\partial t} - \frac{1}{n_f(t,z)}\frac{\partial n_f(t,z)}{\partial t}\right)\frac{\partial P_f}{\partial t} = \frac{e^2 n_f(t,z)}{\varepsilon_0 m_f(t,z)} E = \omega_p(t,z) E \tag{3}$$

Here, $P_f$ is the free electron polarisation, $t$ is time, $\gamma_f$ is the Drude scattering parameter, $e$ is the electronic charge, $\varepsilon_0$ is the vacuum permittivity, and $E$ is the electric field. Both the scattering parameter and plasma frequency are modulated by the time-varying free carrier effective mass $m_f(t,z)$ and the time-varying free carrier density $n_f(t,z)$, which act in opposite directions. By coupling the above equation to Maxwells equations (alongside other auxiliary equations describing the interband process, hot-electron dynamics, and bound electron dynamics), it is possible to model both nonlinearities emerging from a single pump beam. A similar derivation can be completed for the bound oscillators, resulting in:

$$\frac{\partial^2 P_b}{\partial t^2} + \left(\gamma_b - \frac{1}{n_b(t,z)}\frac{\partial n_b(t,z)}{\partial t}\right)\frac{\partial P_b}{\partial t} + \omega_{b,0} P_b = \frac{e^2 n_b(t,z)}{\varepsilon_0 m_{b,0}} E \tag{4}$$

| **Symbol** | **Value** | **Description** |
|---|---|---|
| $n_{f,0}$ | $5 \times 10^{20}$ cm$^{-3}$ | Unpumped number of free carriers. |
| $m_{f,0}$ | $0.4 m_e$ | Unpumped free carrier effective mass. |
| $\gamma_f$ | 15 µm | Free carrier scattering rate. |
| $n_{b,0}$ | $7 \times 10^{21}$ cm$^{-3}$ | Unpumped number of bound carriers. |
| $m_{b,0}$ | $0.8 m_e$ | Bound carrier effective mass. |
| $\gamma_b$ | 1.3 µm | Bound carrier scattering rate. |
| $\omega_{b,0}$ | 280 nm | Bound carrier resonance frequency. |
| $\varepsilon_{bg}$ | 1.27 | Background permittivity. |
| $\Gamma$ | 2.86 ps$^{-1}$ | Relaxation speed for intraband pumping. |
| $\Delta$ | $4 \times 10^{-13} m_e$ kg·cm$^3$/J | Coupling parameter for intraband nonlinearity. |
| $\xi$ | 1.54 ps$^{-1}$ | Relaxation speed for interband pumping. |
| $\Lambda(\lambda_p = 1150$ nm$)$<br>$\Lambda(\lambda_p = 1250$ nm$)$<br>$\Lambda(\lambda_p = 1350$ nm$)$<br>$\Lambda(\lambda_p = 1450$ nm$)$ | 2.1 cm$^3$/J<br>2 cm$^3$/J<br>1.68 cm$^3$/J<br>0.945 cm$^3$/J | Coupling parameter for the interband nonlinearity. |

Table 1: **Fitting parameters for material model.** Symbols, description, and fitted value of model parameters. Here, $m_e$ is the standard electron rest mass.

Here, $P_b$ is the bound electron polarisation, $\gamma_b$ is the bound electron scattering parameter, $\omega_{b,0}$ is the bound electron resonance frequency, $m_{b,0}$ is the bound electron effective mass, and $n_b(t,z)$ is the bound electron density. This last term can be calculated directly from the time dependent free carrier density using the conservation of total carriers $n$, such that $n_b = n - n_f$. The time dependence of $m_f(t,z)$ and $n_f(t,z)$ are driven by hot-electron excitation and carrier elevation processes, respectively. The hot-electron effect can be modelled as follows [15, 29]:

$$\frac{\partial m_f}{\partial t} = \Delta \frac{\partial P_f}{\partial t} E - \Gamma \left( m_f - m_{f,0} \right) \tag{5}$$

Here, $\Delta$ is an absorption cross section describing the transfer of absorbed energy into electronic heating, $\Gamma$ is the hot-electron relaxation parameter, and $m_{f,0}$ is the rest free electron effective mass. The time dependence in the number of free carriers can be described by [31]:

$$\frac{\partial n_f}{\partial t} = \Lambda \frac{\partial P_b}{\partial t} E - \xi (n_f - n_{f,0}) \tag{6}$$

Where $\xi$ is the relaxation rate from the conduction band to the valence band, $n_{f,0}$ is the unpumped number of free carriers, and $\Lambda\left(\lambda_p\right)$ is the absorption cross section describing the transfer of energy absorbed by bound oscillators. This latter parameter is a function of the probe wavelength $\lambda_p$. These auxiliary equations (Eq. 5 and 6) are coupled to the oscillator equations (Eq. 3 and 4), which are subsequently coupled to a second order FDTD solver [31]. Additionally, the material model includes a background permittivity $\varepsilon_{bg}$, which accounts for higher frequency Lorentzian contributions. All parameters were found by fitting simulations to linear and nonlinear data using a differential evolution algorithm, and are tabulated in Table 1 [32].

It is worth noting that all of these fitting parameters were constrained to narrow ranges due to physical restrictions. For example: the number of bound electrons $n_{b,0}$ is set by the density of atoms and does not vary considerably for solids; the background permittivity $\varepsilon_b$ is set by the collective response of high frequency Lorentzian oscillators, and is relatively similar for different solid state materials; the band edge of these materials is known to be in the 250-400 nm wavelength range, fixing the resonance frequency $\omega_{b,0}$ to within a small margin [33]; and the effective mass of free and bound carriers is well known in TCOs from electrical measurements [11]. Finally, several parameters were fixed by fitting the material's linear dispersion ($n_{f,0}$, $m_{f,0}$, $\gamma_f$, $n_{b,0}$, $m_{b,0}$, $\gamma_b$, $\varepsilon_{bg}$, and $\omega_{b,0}$), leaving only a few parameters for fitting the nonlinear responses ($\Gamma$, $\Delta$, $\xi$ and $\Lambda(\lambda)$), as reported in Fig. 2 and Fig. 3.

# Conclusions

In this work, we demonstrated that dual-direction control of the refractive index in TCOs can be achieved using a single-pump excitation scheme, thus overcoming the practical limitations inherent to multi-wavelength approaches. By exploiting the interplay between hot-electron effective-mass nonlinearities and interband nonlinearities, we achieve both positive and negative refractive index modulations of comparable magnitude within the same material platform. This simplified configuration is alternative to synchronized dual-pump wavelength systems while maintaining the ultrafast response and large refractive index modulation characteristic of TCOs compared to other materials. Our results provide a scalable pathway toward engineering arbitrary time-varying optical responses in solid-state systems, opening new possibilities for ultrafast photonic devices and photonic time crystals. In particular, the ability to drive bidirectional index modulation with a single pump beam could enable deeper modulation depths in photonic time crystals, while considerably broadening the intrinsic material bandwidth for efficient on-chip THz generation.